# Tunable thermal conductivity through dual spin-phonon coupling in van der Waals ferromagnetic insulator Cr$_2$Ge$_2$Te$_6$

Zhongbin Wang[1,†], Wenlong Tang[2,†], Simin Pang[3], Hongxing Zhu[1], Renkang Fan[3], Baohai Jia[1], Junxue Li[1,4,5], Ben Xu[2], Jun Zhang[3,6,*], Lin Xie[7] and Jiaqing He[1,4,*]

[1]Department of Physics, State Key Laboratory of Quantum Functional Materials, and Guangdong Basic Research Center of Excellence for Quantum Science, Southern University of Science and Technology, Shenzhen 518055, China

[2]Graduate School of China Academy of Engineering Physics, Beijing 100193, China

[3]State Key Laboratory of Semiconductor Physics and Chip Technologies, Institute of Semiconductors, Chinese Academy of Sciences, Beijing 100083, China

[4]Guangdong Provincial Key Laboratory of Advanced Thermoelectric Materials and Device Physics, Southern University of Science and Technology, Shenzhen 518055, China

[5]Quantum Science Center of Guangdong–Hong Kong–Macao Greater Bay Area (Guangdong), Shenzhen 518045, China

[6]Center of Materials Science and Optoelectronics Engineering, University of Chinese Academy of Sciences, Beijing 100049, China

[7]School of Physical Sciences, Great Bay University, Dongguan 523000, China

†These authors contributed to this work.

*Corresponding author. E-mail: hejq@sustech.edu.cn (J.H.) E-mail: zhangjwill@semi.ac.cn (J.Z.)

The active manipulation of phonon transport remains a central challenge in phononics and spin caloritronics due to the charge-neutral nature of heat carriers. Spin-phonon coupling (SPC) offers a promising route for the dynamic control of heat carriers, yet its progress has been limited due to the lack of a unified framework and suitable material platforms. Here, we report on the magnetic field-tunable phonon transport behavior in the ferromagnetic insulator $Cr_2Ge_2Te_6$. We observed two distinct anomalous regimes at both the high and low fields that were governed by isotropic magnon-phonon hybridization and an anisotropic magnon softening process, respectively. By integrating detailed transport behavior with Brillouin light scattering and ferromagnetic resonance, we uncovered the microscopic origins of these anomalous regimes and demonstrated that both the field magnitude and orientation could act as versatile tuning knobs to manipulate the thermal conductivity. Our findings provide experimental evidence of the SPC effect on phonon transport, demonstrating the dual impact of SPC within a unified system. This work will not only broaden the fundamental understanding of quasiparticle interactions but also establish a viable framework for dynamic phonon engineering. Furthermore, the characteristics of this system highlight the potential for achieving field-tunable phonon transport in similar platforms such as two-dimensional (2D) magnetic materials.

## 1 Introduction

The modulation of phonon transport has attracted increasing attention across a wide range of advanced applications, including solid-state thermal management, thermoelectrics and micro-scale circuitry[1–4]. However, the control of heat currents, which are mainly carried by phonons in semiconductors and insulators, with the same precision as electronic currents, remains a formidable challenge due to the charge-neutral nature of phonons. Spin-phonon coupling (SPC) provides a compelling route to bypass this limitation by bridging phonon transport with the spin degree of freedom. Compared with static phonon engineering approaches such as nanostructuring or alloying[5–7], magnetic tuning via SPC offers the distinct advantage of the dynamic, non-contact, and reversible manipulation of the thermal conductivity $\kappa$.

While the rapid development in spintronics has unveiled numerous phenomena originating from SPC, notably the generation of spin currents by the thermal gradient in the spin Seebeck effect, the reciprocal impact on the phonon counterpart remains overlooked[8–10]. Historically, magnetothermal transport studies have primarily focused on critical behavior, attributing anomalous $\kappa$ behaviors to scattering by critical spin fluctuations or magnetic specific heat near the critical temperature[11–14]. Deep within the magnetically ordered phases, the role of magnons has emerged as a dualistic influence. Magnons participate both as supplementary heat carriers and as potent scatters of phonons via SPC[15–18]. While magnons as carriers follow Bose-Einstein statistics similar to those of phonons, their interaction with phonons underlies spectacular thermal transport anomalies, even extending to the context of the thermal Hall effect and Kitaev physics[19–21]. Although various magnon-phonon scattering processes have been theoretically proposed, a macroscopic thermal transport demonstration remains elusive[22–24]. Specifically, the determination of how these emergent interactions could be harnessed by leveraging the intrinsic magnetic landscape to achieve continuous thermal routing remains an open question.

In this paper we bridge this gap by reporting on highly field-tunable phonon transport behavior in a two-dimensional (2D) van der Waals ferromagnetic (FM) insulator $Cr_2Ge_2Te_6$ (CGT). Taking advantage of the prominent SPC and strong magnetocrystalline anisotropy inherent in 2D magnets, we demonstrated that the external magnetic field acted as a versatile tuning knob for $\kappa$. By mapping a comprehensive magnetothermal phase diagram, we revealed that the applied field selectively activated two distinct phonon scattering mechanisms in different regimes: (i) At high fields, an isotropic suppression of $\kappa$ emerged, which originated from magnon-phonon hybridization (MPH). This created a magnon polaron and selectively reshaped targeted acoustic phonon branches; (ii) At low fields, an anisotropic magnetothermal conductivity (AMTC) is triggered by the dynamic softening of the magnon gap, which regulates the scattering strength of long-wavelength phonons. By integrating macroscopic thermal transport analysis with microscopic Brillouin light scattering (BLS) and ferromagnetic resonance (FMR) spectroscopy, we disentangled the dual impact of SPC on phonon transport within a unified system, expanding the perspective of SPC with respect to phonon transport.

## 2 Results

We employed the standard four-probe method to measure the in-plane (*ab*-plane) longitudinal thermal conductivity, $\kappa_{xx}$, under varying magnetic fields, as illustrated in Fig. 1a. In the figure, the relative orientations (field angle $\theta_H$ and polar angle $\theta_M$) are defined with respect to the *c*-axis (easy axis). For simplicity, we denote this as $\kappa$ hereafter. The Curie temperature ($T_C$) of CGT is about 65 K. Since the thermal transport phenomena of interest occurred at temperatures well below $T_C$, all of the measured $\kappa$ values corresponded to the FM states. Furthermore, the electrical thermal conductivity ($\kappa_e = L\sigma T$) could be neglected due to its large electrical resistivity (beyond measurement limit) at low temperature[25,26]. Fig. 1b displays the temperature dependence of $\kappa$ with various static magnetic fields applied along the *c*-axis of CGT. At zero field, $\kappa(T)$ exhibited typical single-crystal behavior. Upon cooling, $\kappa$ increased due to the freezing out of Umklapp scattering, peaked at approximately 20 K, and subsequently entered the ballistic scattering regime. However, in this regime, the curve deviated from the standard $T^\alpha$ power law expected for a pure phononic system, manifesting a shoulder-like feature, as well as double-peak shape in certain samples (see Fig. S9a). With the application of a magnetic field, $\kappa$ generally increased uniformly due to the suppressed spin fluctuations. However, the $H \parallel ab$ configuration defied this monotonic enhancement at low temperatures and low fields, as shown in Fig. 1c. Specifically, at temperatures below 10 K and fields less than 0.5 T, $\kappa$ dropped by up to ~40% relative to its zero-field value. This counterintuitive reduction strongly implies that the magnetic field did not merely freeze out fluctuations but rather activated new scattering channels. While a typical magnetic resonant scattering model was employed to numerically evaluate the basic magnetic impact on phonon thermal conductivity, this model was insufficient to capture this type of reduction (see Sec 4 in SI). It is noteworthy that the absolute values of our measured $\kappa$ were much higher than previously reported for CGT[25], suggesting a high crystalline quality for our sample. This is crucial for resolving the low-temperature field-sensitive features that remained obscured in earlier studies.

This complex interplay was further elucidated by the isothermal field dependence, $\kappa(H)$, shown in Figs. 1d and 1e. At 30 K, the baseline behavior exhibited a monotonic increase in both configurations, which was consistent with the conventional picture of suppressing spin fluctuations. However, superimposed on this background, two distinct anomalies emerged as the system entered the ballistic scattering regime at low temperatures:

(i) A broad dip at high fields (centered around 4-5 T) was observed for both orientations, which deepened significantly with the decreasing temperature.

(ii) A sharp dip was observed at low fields ($|\mu_0 H| < 0.5$ T), exclusively for the $H \parallel ab$ configuration, whereas $\kappa$ for $H \parallel c$ remained essentially flat in this regime.

The coexistence of these isotropic and anisotropic features across different field regimes highlighted the exceptional sensitivity of phonons to the magnetic field in CGT. Furthermore, this pointed to two distinct magnon-phonon scattering processes that could alter the thermal transport: magnon-phonon hybridization (MPH) and the magnon-number non-conserving confluence process[24].

**High-field magnon-phonon hybridization**

To elucidate the microscopic origin of the high-field suppression, we analyzed the relative change in the thermal conductivity, $\Delta\kappa/\kappa_0 = [\kappa(H) - \kappa(0)]/\kappa(0) \times 100\%$, in Fig. 2a. Given the symmetry with respect to $\pm H$, a half-part from each orientation was compared. A prominent dip emerged in both orientations within the same field regime, suggesting an intrinsic high-field isotropy that strongly contrasted with the low-field anisotropy. The overall response was dominated by a monotonic background arising from the suppression of spin fluctuations. To extract the dip feature, we approximated this background using an empirical exponential function (dashed lines in Fig. 2b upper panel), excluding the resonance region during fitting. Subtracting this baseline yielded the residual dip, $(\Delta\kappa/\kappa_0)_{dip}$, which exhibited a pronounced Lorentzian profile centered at $H_{dip} \approx 4.5\text{-}5.5$ T (Fig. 2b lower panel). Fig. 2c summarizes the temperature evolution of the resonance characteristics. The dip amplitude decayed rapidly with the increase in temperature, while the dip field $H_{dip}$ remained stable for both orientations (Fig. 2c inset). The constant offset between $H \parallel ab$ and $H \parallel c$ (~0.5 T) arose from the magnetic anisotropy gap in the magnon dispersion[27]. The temperature independent $H_{dip}$ implied a spectral-level origin. The specific scattering process was only activated when the field tuned the excitation spectra into a resonance condition.

Consequently, we demonstrated that this field-induced suppression was due to MPH. As illustrated by the calculated dispersion in Fig. 2d, the applied field shifted the magnon branch upward via the Zeeman effect ($\Delta E \propto g\mu_B H$), driving it to intersect with the acoustic phonon branches[28]. This interaction formed an anti-crossing band, creating a hybrid quasiparticle gap that scattered the targeted phonon mode[22,29]. The resonant scattering was maximized when the magnon and phonon dispersions became tangential, because this maximized the overlapping phase space for the magnon polaron. The magnon polaron model consistently explains the temperature-independent dip field and the decay of the dip amplitude. The former reflects the rigid intrinsic dispersions of magnon and phonon branches, whereas the latter indicates that the magnon-phonon coherence was destroyed by the enhanced phonon-phonon scattering process as the temperature increased[30].

Specifically, while the model predicted intersections with both the transverse (TA2, ~2 T) and longitudinal (LA, ~5.5 T) branches, our experimental results predominantly captured the LA crossing (The out-of-plane mode TA1 had the lowest group velocity and a rather small contribution to the in-plane $\kappa$). This observation suggested a strong mode-selective effect. In contrast to the spin Seebeck effect, for which multiple magnon polaron crossings enhanced the spin current signal[29,31], not every avoided crossing yielded a macroscopic impact on $\kappa$. The prominent LA signature implied that these longitudinal phonons either dominated the phonon transport due to the large group velocity or possessed stronger intrinsic coupling strengths than other modes.

**Low-field anisotropic magnetothermal conductivity**

In contrast to the isotropic high-field resonance, the low-field regime for $H \parallel ab$ exhibited a sharp, field-symmetric suppression confined within the saturation field ($H_s$ = 0.5 T). As illustrated in Fig. 3a, the $\Delta\kappa/\kappa_0$ dropped to a pronounced minimum at ~ 0.3 T before recovering toward $H_s$, stabilizing at a value distinct from its zero-field value. Given the nearly field-independent behavior below 0.3 T for $H \parallel c$, the strict confinement within the $H_s$ boundary strongly linked the anomaly to magnetocrystalline anisotropy (Fig. S3) [32,33]. However, the non-monotonic nature of the dip implied that the phonon transport was not a simple linear function of the magnetization orientation. For $H < H_s$, the magnetization rotated continuously from the out-of-plane to the in-plane direction, parameterized by the polar angle $\sin\theta_M = H/H_s$. Inspired by the phenomenology of anisotropic magnetoresistance (AMR), we proposed that the AMTC was primarily governed by the angle $\phi$ between the magnetization and the heat current ($\phi = 90° - \theta_M$). A Fourier expansion was used to capture this orientation dependence[34,35]:

$$\Delta\kappa/\kappa_0 = \sum_{n=0}^{N} C_{2n} \cos(2n\phi) \qquad (1)$$

where $C_{2n}$ are the fitted parameters. The data is well described by an $n$ = 3 expansion (dashed lines in Fig. 3a; see Fig. S6 for $n$ = 2 comparison). To validate this orientation dependent model, we performed independent angular dependent measurements by rotating the sample within a fixed magnetic field direction $\theta_H$ (Fig. 1e). This effectively tuned the in-plane magnetization component $M_x$ while maintaining a constant field magnitude. As shown in Fig. 3b, the $\Delta\kappa/\kappa_0$ data were extracted from angular rotation measurements at $H \approx \pm 0.5$ T, where $M$ fully aligned with $H$ at various angles ($\theta_M = \theta_H$) and without the introduction of an additional Zeeman effect, which qualitatively agreed with the field-sweep data (dashed lines) transformed via $\sin\theta_M = H/H_s$ from Fig. 3a. This consistency between the distinct experimental approaches provided robust evidence that the AMTC was fundamentally governed by the magnetization orientation. Despite sharing a functional similarity with AMR, the physical origin of this AMTC was fundamentally distinct. We attributed the necessity of the high-order expansion terms to the polar angle dependence of the uniaxial magnetocrystalline anisotropy energy (MAE)[36]:

$$\frac{E_a}{V} = K_1 \sin^2\theta + K_2 \sin^4\theta + K_3 \sin^6\theta \qquad (2)$$

While the anisotropy constant $K_u$ is typically dominated by the first term, the deviation from Callen-Callen power law in CGT indicates the higher-order term contribution arising from the strong spin-orbit coupling induced by the heavy ligand Te $p$ orbital[37,38]. The exceptional sensitivity of phonon transport thus enables us to capture the influence of often-overlooked high-order MAE terms on AMTC behavior.

The MAE-driven angular dependence indicates that the magnon dispersion underwent a fundamental softening process. In ferromagnets with perpendicular magnetic anisotropy (PMA), the magnon spectrum possesses a finite zero-field gap ($\Delta_\Gamma \neq 0$). When a field orthogonal to the easy axis is applied, the anisotropy energy must be overcome, leading to a soft magnon behavior[39]. The magnon frequency at the $\Gamma$ point could be directly probed with FMR, as shown in Fig. 3c. Fitting the high-field data with the Kittel formula yielded a critical field of ~0.3 T at low temperatures, below which the theoretical magnon frequency follows a characteristic square-root relationship (dashed red line)[37]. Because a low-frequency magnon could scatter long-wavelength phonons, which were the dominant low-temperature heat carriers, the magnon population was the key factor governing the reduction in $\kappa$. This occurred via a magnon-number non-conserving confluence process, which is forbidden for phonon energies below twice the magnon gap[24,40]. As this gap closed, the Bose-Einstein distribution dictated an exponential surge in the thermal magnon population, thereby intensifying phonon scattering and suppressing $\kappa$. As the field approached 0.3 T, the gap closed completely, which maximized the magnon population and allowed all of the phonons in momentum space to be scattered. Beyond 0.3 T, the gap reopened, and the magnon population decreased until the saturation field at 0.5 T, after which the magnon dispersion followed the typical linear Zeeman shift. This dynamic gap evolution perfectly explained the tuning point and the sharp confinement of the AMTC behavior. Notably, in the positive deviation in the low-field FMR data suggested the presence of multi-domain, as reported in previous studies[41–43]. However, the lack of thermal hysteresis during field sweeping (i.e., $\kappa(H) = \kappa(-H)$) effectively ruled out static domain wall scattering as the primary mechanism for this AMTC.

To map the low-frequency magnetic excitations near the $\Gamma$ point, we further performed field-dependent BLS measurements (Fig. 3d). Below ~ 0.3 T, two distinct magnon branches emerged: sharp peaks, denoted as acoustic modes (AM), which shifted approximately linearly with $H$, and broad peaks, denoted as optical modes (OM), which underwent frequency softening with $H$. While typical for antiferromagnets, such low-frequency OMs can also exist in ferromagnets due to the confined spin precession within nanoscale domain walls[44]. Above ~ 0.5 T, where $M$ was fully aligned with $H \parallel ab$, the frequencies of the uniform ferromagnetic modes (FM) increased linearly with $H$ (Fig. S7a). The extracted intensities (Fig. 3e) showed that the AM intensity peaked sharply at 0.3 T. This confirmed the maximized magnon population available for phonon scattering, perfectly corroborating the FMR softening results. At the same time, the broad linewidth of the OM implied a short quasiparticle lifetime ($\Delta\nu \approx 1/(2\pi\tau)$), hinting at strong magnon-phonon coupling. The OM softening and broadened full widths at half maximum (FWHM) toward 0.3 T (Fig. 3f) indicated enhanced coupling. This coupling weakened as the field increased further, directly mirroring the suppression and subsequent recovery of $\kappa$. In contrast, the out-of-plane ($H \parallel c$) BLS spectra (Fig. S7) showed only marginal frequency shifts below 0.3 T, with intensities roughly an order of magnitude lower. This naturally explains the negligible effect on $\kappa$ for the $H \parallel c$ configuration.

**Tunable Magnetothermal Transport Phase Diagram**

To visualize the comprehensive impact of SPC, we constructed a continuous $H$-$\theta$ magnetothermal transport phase diagram (Fig. 4a) by interpolating discrete $\kappa(H)$ datasets (Fig. S8). This contour map of $\Delta\kappa/\kappa_0$ clearly delineates two contrasting regimes that were dictated by competing physical processes. In the red regions, the suppression of spin fluctuations reduced the background magnetic scattering, manifesting as a net enhancement of $\kappa$ ($\Delta\kappa/\kappa_0 > 0$). Conversely, the blue regions signify an anomalous suppression ($\Delta\kappa/\kappa_0 < 0$), where the emergent SPC scattering completely overwhelmed this background enhancement. The interplay between these competing mechanisms dictated the intricate boundaries of the suppression valleys across the $H$-$\theta$ parameter space.

The most striking feature of the contour map is the evolution of the suppression zones as the field tilted away from the hard $ab$-plane (90°) toward the easy $c$-axis (0°). At $\theta = 90°$, the sharp AMTC dip is highly localized around 0.3 T, forming a deep, narrow blue pocket. (The high-field MPH suppression at this angle was too subtle to dominate the global color scale, though clearly visible in the line profiles of Fig. S8). As the field tilted toward the $c$-axis (decreasing $\theta$), this low-field suppression region broadened significantly and shifted to higher fields, directly mapping the angular dependence of the magnon gap softening. Notably, at intermediate angles, the previously distinct AMTC and MPH regimes blurred and merged into a wide, continuous suppression valley. This revealed that these two distinct SPC mechanisms could effectively overlap in the energy space, driving a cooperative suppression of phonon transport over an extended field range.

Finally, this highly structured phase diagram was extremely sensitive to temperature. As $T$ increased, the blue suppression valleys rapidly shrank and lost depth (Figs. 4b, c). This thermal quenching occurred because the enhanced phonon-phonon scattering at higher temperatures effectively destroyed the magnon-phonon coherence required for these SPC signatures. By 30 K, the intricate non-monotonic features had vanished entirely, leaving only the featureless, monotonic background characteristic of the Umklapp scattering regime. Ultimately, this global mapping demonstrated that the synergistic tuning of the magnetic field magnitude and orientation provided a robust, nonlinear knob for continuously modulating the macroscopic phonon transport.

## 3 Conclusion

In summary, we have presented a comprehensive analysis of the field-tunable phonon transport in CGT. By integrating magnetothermal measurements with spectroscopy and theoretical modeling, we unraveled a dual SPC mechanism on thermal transport. In the high-field regime, the resonant scattering by magnon polarons provided a direct bulk manifestation of the reciprocal interaction foundational to spin caloritronics[45]. At the same time, in the low-field regime, the discovery of the orientation dependent AMTC revealed the potent role of MAE and higher-order terms in modulating heat flow.

Beyond CGT, this magnetothermal transport framework offers a versatile template for exploring SPC across a diverse spectrum of magnetic materials. Crucially, our findings demonstrate thermal conductivity not merely as a transport coefficient, but as a highly sensitive spectroscopic probe for emergent quasiparticle dynamics. From an application perspective, the non-contact thermal modulation that was achieved in this research, particularly at practically accessible low fields, will position 2D vdW magnets as promising candidates for active thermal management. Ultimately, this study will enable novel concepts in phononic device design, paving the way for further advancements in phonon engineering and spin-caloritronics.

**Methods**

**Material synthesis and transport measurement**

Single Crystal Growth: High-quality CGT single crystals were synthesized via the self-flux method with a precursor ratio of 1:3:18 (Cr:Ge:Te), following established protocols reported[26]. The resulting crystals exhibited pristine surfaces and were cut into rectangular bars with dimensions up to ~8×3×0.2 mm for subsequent transport measurements.

Thermal Transport Characterization: The longitudinal thermal conductivity ($\kappa_{xx}$) was measured using the thermal transport option (TTO) in a Physical Property Measurement System (PPMS, Quantum Design). To ensure high measurement precision and eliminate contact resistance effects, a standard four-probe configuration was employed across all temperature and field ranges. The thermal current was applied along the in-plane direction of the crystals. Optical images for two configuration were captured to verify the precise alignment and geometry, as documented in Supplementary Fig. S1. To measure the angular dependence of magnetothermal conductivity, as the sample was mounted with copper bar, it was fixed by bending the copper bar to several angles for each time. The relative angle $\theta$ between the field and the crystallographic c-axis was carefully calibrated.

**Ferromagnetic resonance (FMR) measurements**

The ferromagnetic resonance (FMR) measurements were performed in a Physical Property Measurement System (PPMS, Quantum Design) with a 9 T superconducting magnet at 3-30 K. During the FMR measurement, a CGT crystal was mounted on a custom FMR probe equipped with a coplanar waveguide (CPW) and a Helmholtz coil for field modulation. Field modulation was achieved by driving an alternating current (50 mA, 5 kHz) from a current source (Keithley 6221) to the integrated Helmholtz coils. An external magnetic field was applied perpendicularly to the [0001]-axis of the CGT flake. Continuous microwaves (up to 40 GHz) from a microwave source (Rohde and Schwarz SMA100B) were delivered to the CPW. The transmitted microwave signal was rectified by a diode detector and measured with a lock-in amplifier at 5 kHz (Stanford Research Systems SR830).

**Brillouin light scattering**

Brillouin light scattering (BLS) spectroscopy was performed at 4 K in a backscattering geometry using a commercial system (JRS Scientific Instruments) that combined a confocal microscope (CM-1) with a high-contrast (3+3)-pass tandem Fabry–Pérot interferometer (TFP-2 HC). The sample was mounted in a closed-cycle cryostat (attoDRY 1000). Two different cryogenic-compatible objectives were used for the distinct magneto-optical geometries: A 50× objective (numerical aperture NA = 0.68) for the Voigt geometry, and a 100× objective (NA = 0.82) for the Faraday geometry. A 532 nm single-longitudinal-mode laser, focused at normal incidence with a power of ~1.8 mW, served as the excitation source. All of the spectra were acquired with VH polarization geometry. Magneto-BLS measurements were conducted in two field orientations: the Voigt geometry (in-plane field, up to 9 T, perpendicular to the light wavevector) and the Faraday geometry (out-of-plane field, up to 9 T, parallel to the light wavevector), which enabled mode-selective probing.

**Acknowledgments**

The work was supported by the National Key Research and Development Program of China (Grant Nos. 2024YFA1210400, and 2025YFA1411302), the National Natural Science Foundation of China (Grant Nos. 12434001, 52461160258, U25A20239, 12534003, and 12374111), Guangdong Provincial Key Laboratory of Advanced Thermoelectric Materials and Device Physics(Grant No. 2024B1212010001), Guangdong Basic and Applied Basic Research Foundation (Nos. 2026B0303000004, and 2024JC08A027), Guangdong Provincial Quantum Science Strategic Initiative (No. GDZX2301004), the Outstanding Talents Training Fund in Shenzhen(Grant No. 202108), the Science, Technology and Innovation Commission of Shenzhen Municipality(Grant No. ZDSYS20141118160434515). J. Z. acknowledges the funding support from the National Natural Science Foundation of China (12525405) and the CAS Project for Young Scientists in Basic Research (YSBR-120).

**Author Contributions**

Z.W. and J.H. conceived the research and designed the experiments. Z.W. synthesized the single crystals and performed the thermal transport measurements. W.T. provided the spectral calculations. S.P. and R.F. conducted the BLS measurements, while H.Z. performed the FMR measurements. Z.W., S.P., W.T., L.X., J.L., and J.Z. collaborated on data analysis and interpretation. Z.W. and J.H. drafted the manuscript with contributions from all authors. L.X., J.L., B.J., and J.H. contributed to the critical revision of the manuscript.

**Competing interests**

The authors declare no competing interests.

**Data and materials availability**

All data needed to evaluate the conclusions in the paper are present in the paper and/or the Supplementary Information.

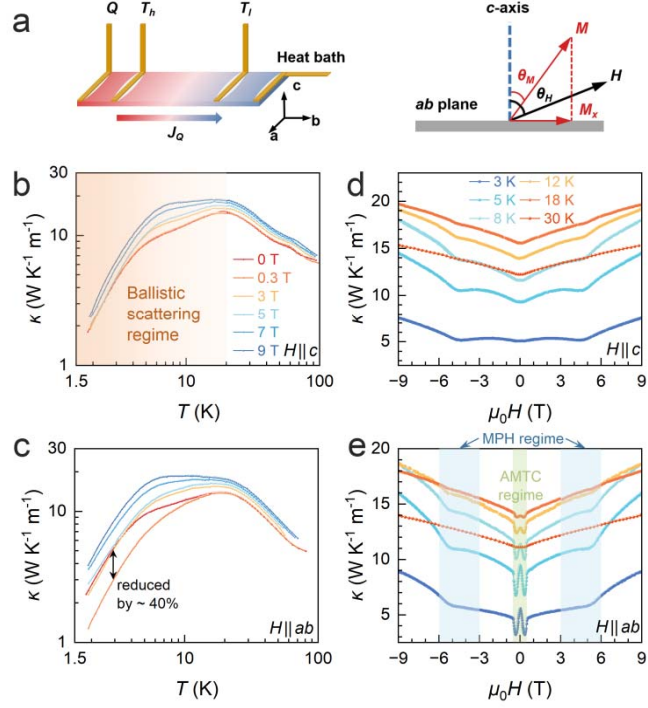

**Fig. 1: Magnetothermal transport in the dual magnon-phonon coupling regime of $Cr_2Ge_2Te_6$ (CGT).**
(a) Schematic of the four-probe thermal transport measurement configuration, illustrating the geometric relationship between the applied magnetic field ($H$), the magnetization ($M$), and the heat current ($J_Q$). (b, c) Temperature dependence of thermal conductivity $\kappa$, with representative magnetic fields applied along the $c$-axis (easy axis) (b) and within the $ab$-plane (hard plane) (c). The shaded area highlights the ballistic scattering regime. (d, e) Isothermal magnetothermal conductivity $\kappa$ as a function of the magnetic field applied along the $c$-axis (easy axis) (d) and in the $ab$-plane (hard plane) (e) at various temperatures. The shaded areas in (e) highlight the sharp anisotropic magnetothermal conductivity (AMTC) regime and the broad magnon-phonon hybridization (MPH) regime.

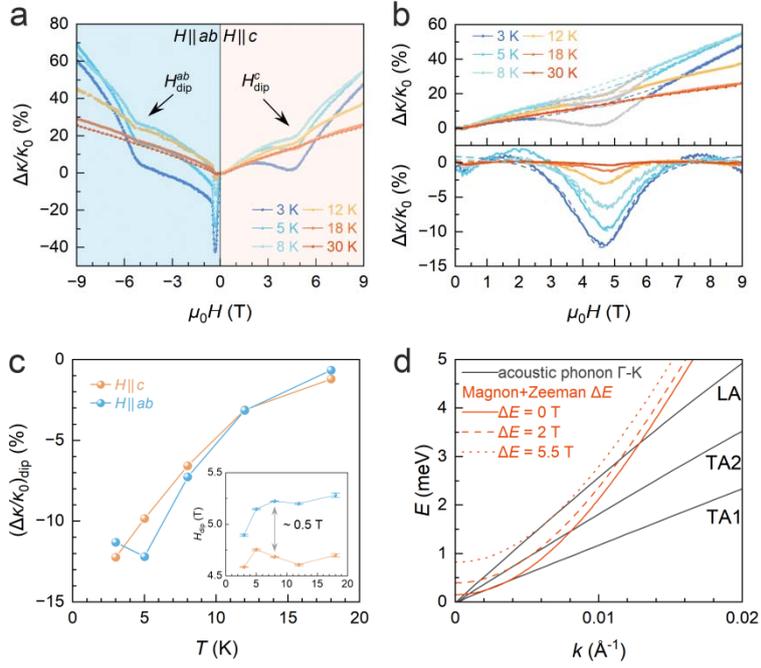

**Fig.2: High-field resonance dip originating from magnon-phonon hybridization.**

(a) Magnetic field dependence of the relative change in the thermal conductivity $\Delta \kappa / \kappa_0 = [\kappa(H) - \kappa(0)]/\kappa(0) \times 100\%$, for $H \parallel ab$ (left) and $H \parallel c$ (right). (b) Upper panel: Representative baseline extraction process for $H \parallel c$. The monotonic baselines (dashed lines) were fitted using an empirical exponential function, excluding the resonance region (3-6 T). Lower panel: Extracted residual dips $(\Delta\kappa/\kappa_0)_{\text{dip}}$, obtained by subtracting the baseline and subsequently fitted with Lorentzian profiles (dashed lines). (c) Temperature evolution of the dip amplitudes for both $H \parallel ab$ and $H \parallel c$ configurations. The inset shows that the dip positions remained nearly constant with the temperature for a magnetic filed near ~ 5 T. The 0.5 T difference between the two configurations corresponds to the magnetic anisotropic energy (MAE) in CGT. (d) Calculated phonon (gray lines) and magnon (orange lines) dispersion relationships along the Γ-K direction at 0 T (solid), 2 T (dashed), and 5.5 T (dotted), illustrating the field induced magnon-phonon hybridization.

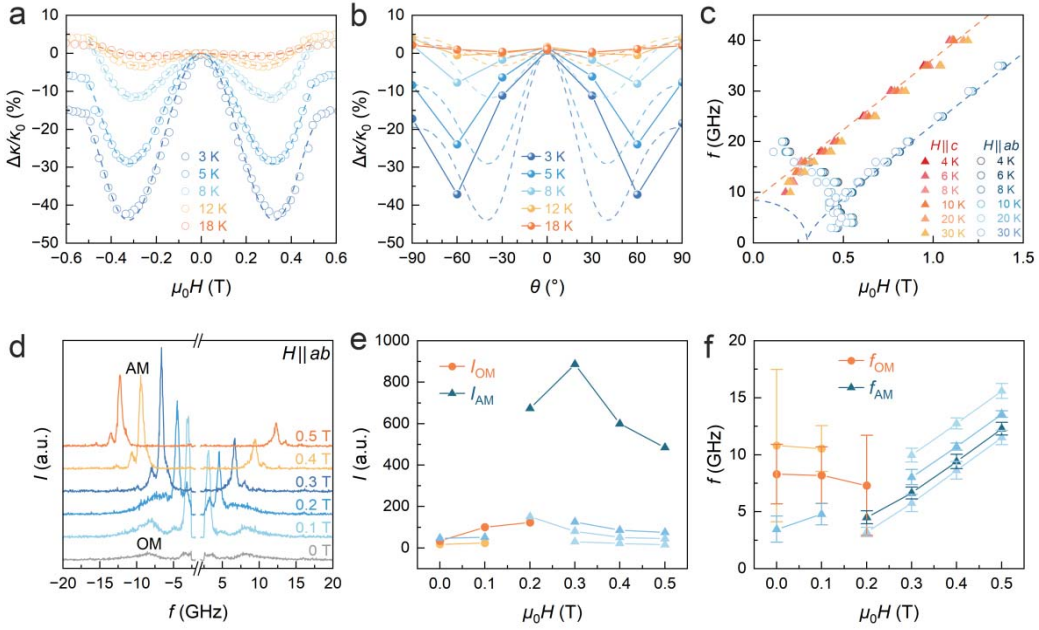

**Fig. 3: Low-field anisotropic magnetothermal conductivity originating from magnetic anisotropy energy.**

(a) Low-field relative change in thermal conductivity $\Delta\kappa/\kappa_0$, for $H \parallel ab$. The dashed lines represent the fits using Equation (1), as described in the main text. (b) Angular dependence of $\Delta\kappa/\kappa_0$ measured at the polar angles $\theta = 0°, 30°, 60°$, and $90°$ for a constant applied field of $\mu_0 H = \pm 0.5$ T. The dashed lines show the transformed angular dependence derived from (a) using the relationship $\sin\theta_M = H/H_s$. (c) Ferromagnetic resonance frequency evolution for $H \parallel c$ (solid triangles) and $H \parallel ab$ (open circles). The dashed lines represent the single-domain Kittel model predictions. (d) Brillouin light scattering spectra for varying magnetic fields ($H \parallel ab$) from 0 T to 0.5 T. The sharp and broad peaks are denoted as acoustic modes (AM) and optical modes (OM), respectively. (e) Magnetic field dependence of the intensities for the AM (triangles) and OM (circles). The brightest colors denote the strongest representative peaks for AM and OM. (f) Magnetic field dependence of the frequency and full-widths at half-maximum (denoted by the error bar) for AM and OM.

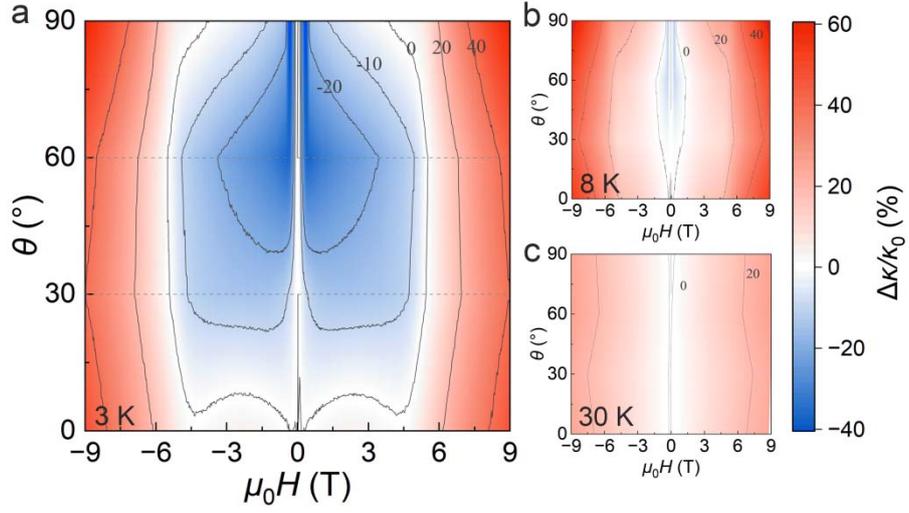

**Fig. 4: Magnetothermal transport phase diagram Δκ(H, θ) at low temperatures.**

(a) Contour map of the relative change in the thermal conductivity, $\Delta\kappa/\kappa_0$, as a function of magnetic field magnitude $\mu_0 H$ and the orientation $\theta$ at $T$ = 3 K. The red areas represent the thermal conductivity enhancement ($\Delta\kappa/\kappa_0 > 0$) for a high field due to the suppression of the spin fluctuation, while the blue areas indicate the anomalous suppression ($\Delta\kappa/\kappa_0 < 0$) for a low field driven by SPC. The solid lines represent constant $\Delta\kappa/\kappa_0$ contours. (b, c) Temperature evolutions of the transport phase diagram at 8 K and 30 K, illustrating the thermal quenching of the SPC features. As the temperature increased, the low-field suppression region (blue regions) rapidly shrank. By 30 K, the field-induced suppression of $\kappa$ had entirely vanished, indicating that the SPC effect was completely overwhelmed by the dominant phonon-phonon Umklapp scattering at elevated temperatures.